\documentclass{article}
\usepackage[dvips]{graphicx}
\usepackage{natbib}
\bibpunct{(}{)}{,}{a}{}{,} %

\begin{document}
\begin{center}
{\bf \LARGE Supernovae and Cosmology\footnote{appeared in General
Relativity and Gravitation, 40, 221 (2008) --
http://www.springerlink.com/content/0xr331rk310lqx67/}}

{\bf \normalsize
Bruno Leibundgut\\
European Southern Observatory\\
Karl-Schwarzschild-Strasse 2, D-85748 Garching\\
Germany \\
Email: bleibundgut@eso.org}
\end{center}

\begin{abstract}

The extreme luminosity and their fairly unique temporal behaviour have
made supernovae a superb tool to measure distances in the universe.
As complex astrophysical events they provide interesting insights
into explosion physics, explosive nucleosynthesis, hydrodynamics of
the explosion and radiation transport. They are an end product of
stellar evolution and provide clues to the stellar composition. Since
they can be observed at large distances they have become critical
probes to further explore astrophysical effects, like dust properties
in external galaxies and the star formation history of galaxies. Some
of the astrophysics interferes with the cosmological applications of
supernovae. The local velocity field, distorted by the gravitational
attraction of the local large scale structure, and the reddening law
appear at the moment the major limitations in the accuracy with which
cosmological parameters can be determined. These absorption effects
can introduce a secondary bias into the observations of the distant
supernovae, which needs to be carefully evaluated. Supernovae have
been used for the measurement of the Hubble constant, i.e. the current
expansion rate of the universe, and the accelerated cosmic expansion
directly inferred from the apparent faintness of the distant
supernovae. 

\end{abstract}


\section{Introduction}
\label{sec:intro}

The energetic display of a supernova marks the transition from a bound
star to the recycling of material into the gas pool of a galaxy or
beyond. The progenitor star at explosion could still have an active
nuclear furnace operating or could be a degenerate end product of
stellar evolution. The corresponding results also take different
forms: a compact ``stellar'' remnant, a neutron star or a black hole
as the result of a collapse of the stellar core,
or no compact remnant, when the star is incinerated by a nuclear
explosion. In all cases, the expelled material will
interact with its environment and produce a supernova remnant. One of
the main topic of interest is how the different physical processes
lead to the observed displays.  As further exposed in the following,
some of the uncertainties in our understanding of the supernova
physics limits their use in cosmological applications. 

Supernovae shaped today's universe in many different ways. They are
the main mechanism to create heavy elements, especially the ones only
created in explosive nucleosynthesis. They are also responsible for
the return of these newly created elements into the baryonic cycle of
dust, gas and stars. The energy input into the interstellar material
can be so significant that star formation can be triggered or
suppressed. For smaller galaxies, supernovae most likely shape their
appearances. Cosmic ray acceleration is most probably done in the
shock of supernova remnants and the collapse of massive stellar cores
are the main source of neutrinos beyond the Big Bang. 

Supernovae appear in very different displays. In fact, a clear
definition of a supernova does not exist. There is a classification
scheme, which dates back to Walter Baade, Fritz Zwicky and Robert
Minkowski \citep{baa34,min41,min64}. For a modern version with detailed
definitions see \citet{fil97}. A supernova in the following will be the
event when a star ejects most of its material in a violent explosion and
ceases to exist as a stellar entity. Note that this is a physical description,
while the observations we obtain are often not able to definitely
ascertain that the above condition is fulfilled. Nevertheless, a
supernova by definition cannot be recurrent. It marks the end of the
existence of a star as an individual object. One should note that this
definition includes $\gamma-$ray bursts together with the more
traditional supernova classes. 

Due to their luminosity supernovae have been a favourite for
cosmological applications. They are also markers
of star formation and could be amongst the earliest objects we may be
able to observe in the early universe. 

This article first presents a brief history of supernovae. It will then
comment on the current classification scheme and its use to understand
the explosion physics and the radiation hydrodynamics, which takes place
in these explosions. Supernovae as cosmological distance indicators will
be examined first before we will move on to a discussion of the Hubble
constant and the expansion history of the universe as derived from
supernovae. The latter is currently concentrated on Type Ia supernovae
(SNe~Ia hereafter), which have been the most successful in measuring
distances half way across the universe. 

The literature on supernovae and their cosmological applications has
literally exploded in the past decade. There are the classic papers,
which will be mentioned in this review, but also many associated
interpretations. Overviews have been presented in recent monographs on
supernovae and gamma-ray bursts
\citep[e.g.][]{nie00,hil03,wei03,hof04,mar05,tur05}. Supernova physics is
reviewed in \citet{fil97,hil00,lei00} and \citet{woo07a}.
Several reviews of the supernova cosmology have been published as well
\citep{bra98,rie00,lei01a,per03}.

\subsection{Some early history}
The appearance of new stars, "stellae novae" from their Latin
designation, has always intrigued astronomers as documented in the
ancient Chinese and Korean records (see \citet{cla77} and \citet{mur78}
for reviews of the historic supernovae in the Milky Way observed over
the last two millennia). 

The first to suggest that there are two classes of novae was 
\citet{lun25}, who proposed an 'upper class' about 10 magnitudes
brighter than the 'lower class' of novae. The latter would correspond to
the well known Galactic novae. He based his proposal mostly on the
observation of the (super)nova 'S~Andromeda' observed in 1885 (designated
SN~1885A in modern nomenclature), which appeared that much brighter
than a sample of about two dozen regular novae in the Andromeda
galaxy. Lundmark later seemingly was the first to suggest the name
'super-nova' \citep{lun32}. 

It was Walter Baade who made the connection between the historical
supernovae and the observed emission nebulae at their positions, thus
identifying the remnants of the explosions. The most prominent object is
of course the Crab Nebular (Messier 1), the leftover from the supernova
in 1054 \citep{baa42,may42}. With extensive observations
of bright supernovae \citet{min41} introduced two subclasses. 
\citet{zwi65} refined the classification scheme for supernovae further.
However, for several decades only two main classes were maintained until
in the early 1980s it became clear that at least one further subclass
needed to be added. The classification scheme has now expanded again
with the introduction of
several subclasses to further distinguish between different
observed displays. Some proposals mix spectroscopic definitions with the
light curve appearance, while others even introduced theoretical
arguments into the classification. The reason for a classification scheme
should remain simple and it should not be mixed with 
theoretical ideas. While different behaviour clearly indicates different
physics, the classification as used in the past was primarily to quickly plan
observing strategies and give an indication what type of event was
observed. This still is often the case for the projects, which make
use of SNe~Ia for cosmology, as the spectroscopy time
needs to be used as efficiently as possible.

\section{Supernova classification}
\label{sec:class}

\begin{figure}
\centering
\includegraphics[width=10.5cm]{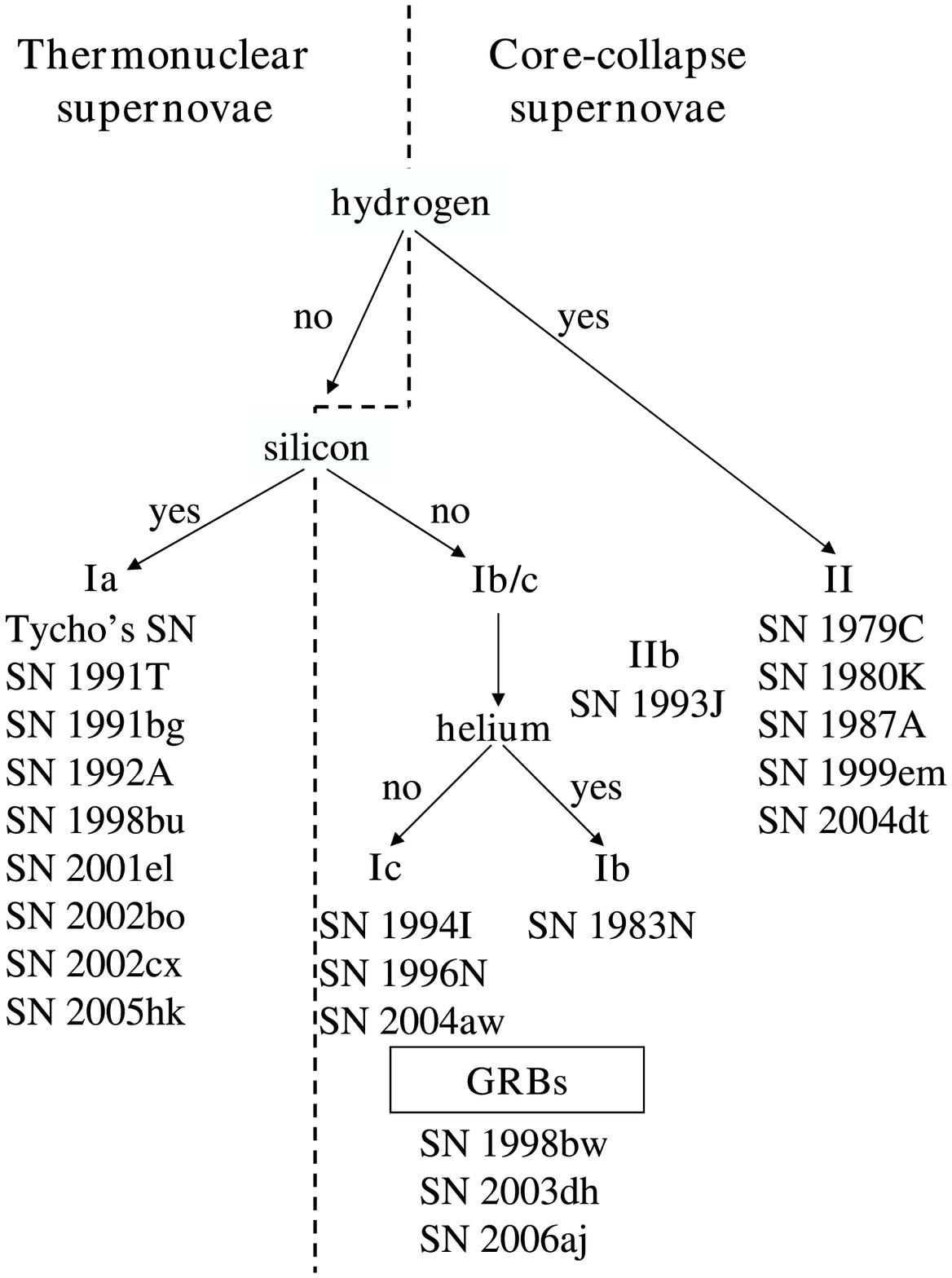}
\caption{Classification scheme for supernovae. The presence or absence
of specific absorption features in the maximum-light spectrum is used
to separate the supernovae into different classes. The SNe~Ia are
the only ones which are thought to come from the thermonuclear
explosion of a white dwarf. All others are powered by the core
collapse in a massive star. Prominent examples of the various classes
are indicated.}
\label{class}
\end{figure}

The modern classification of supernovae is based on the spectroscopy
at maximum light (e.g. \citealt{fil97,tur03} - see also
Fig.~\ref{class}). The distinction is done through the presence (or
absence) of hydrogen lines in the optical spectra near maximum
brightness leading to the classes of Type II supernovae (or Type I
supernovae). The hydrogen-deficient supernovae are further subdivided
into groups which display prominent absorption near 6150\AA\
attributed to a transition in singly ionised silicon (Si~II in
astronomical notation) for the Type Ia supernovae and others which
show sodium and oxygen absorption lines, designated Type Ib/c
supernovae (Fig.~\ref{class}). The separation of these two subclasses
happened during the early 1980s, when it became clear that there was a
subset of Type~I supernovae that showed very red colours, a spectral
evolution, which appeared accelerated, and showed lines of
intermediate elements at late phases \citep{whe85,uom85,pan86,fil86}.
The presence/absence of helium lines is used as a separation into the
Type Ib/Type Ic supernovae, respectively. The exact physical
interpretation of this separation remains relatively weak.  An
evolutionary sequence for the separation of these core-collapse
supernovae has been proposed, in which the appearance is determined by
the amount of hydrogen envelope remaining on the star at the time of
explosion. Regular stars with a thick hydrogen layer would explode as
SNe~II, while the ones which lost this hydrogen layer, e.g. due to a
strong stellar wind or interaction with a binary companion, would
become SNe~Ib. Should the helium layer be eroded as well, then a SN~Ic
is observed. Moreover, there is one 'cross-over' class of Type IIb
supernovae, with the prominent example of SN~1993J. These events typically
start out as hydrogen-displaying supernovae (hence SN~II) before the
hydrogen lines disappear and the objects start to resemble Type Ib/c
supernovae. They represent the major link showing that the SNe~Ib/c
are core-collapse supernovae. Figure~\ref{class} also lists some
prominent examples for each SN class.

A physical picture for this classification scheme has emerged. The
Type Ia supernovae are coming from thermonuclear explosions of stars,
which have shed hydrogen and helium during their progenitor evolution.
Hence no traces of these elements are observed in these explosions.
All other supernovae most likely come from the core collapse in
massive stars or in some cases more exotic phenomena, like pair
instability \citep[e.g.][]{heg03}. The signature for these events are
their oxygen and calcium rich spectra at late phases. 

It is notable that gravity is the ultimate reason for both types of
explosions. In the cores of massive stars the hydrostatic equilibrium
is maintained by burning to higher and higher elements at increasing
temperatures. By the time the core has burnt its fuel to iron no
further exothermic reactions are possible and the stellar core
collapses under the weight of the outer layers of the star. The
collapse is only stopped when the material reaches nuclear densities
where electrons and protons merge and create neutrons. At this stage
the proto-neutron star provides a hard surface. The neutrinos created
in this process emerge mostly without interacting, but even a tiny
amount of energy deposited by the neutrinos in the envelope can turn
the implosion into an explosion.  The exact mechanism has not been
fully explored, but at least small stars (8 -- 10 M$_\odot$) can now
be made to explode moderately by the modellers
\citep{kit06}. 

Hypernovae have been added to the list of supernovae and they
represent the high energy end (at least in their kinematics) with the
large expansion velocities observed in these objects. The connection
of gamma--ray burst with supernovae has now been generally accepted
with the observations of SN~2003dh/GRB030329
\citep{sta03,mat03,hjo03}.  It should be noted that already
SN~1998bw/GRB980425 showed all the signatures of a supernova
\citep{gal98,pat01}. Hypernovae are characterised by the
absence of hydrogen and helium and very high expansion velocities
observed in their spectra \citep{maz02,maz03,woo07a}. In some cases no
gamma-ray burst is observed, like for SN~2002ap. The amount of nickel
synthesised in these explosions is substantial \citep[up to about 0.5
$M_{\odot}$;][]{sol02}.  The kinetic energies inferred from the
line widths are also substantially larger than the ones of regular
SNe~II. In many aspects they appear to be similar to the SNe~Ib/c with
high kinetic energy.

In the case of the thermonuclear supernovae the electron-degenerate
white dwarf has to cope with an increasing amount of material piled onto
it by a companion star, and hence increases the pressure and temperature
in the core. Again, it is the gravitational force which sets off the
explosion, in this case the explosive carbon and oxygen burning, which
disrupts the star. A comparison of the binding energy of a neutron star
or the binding energy of a solar mass of iron give a clear
indication of how much energy is released in these explosions. 

Some objects cannot be clearly classified into one or the other class.
Prime examples are SN~2002ic and SN~2006gy, which both have been
interpreted as possibly a thermonuclear explosion or a core-collapse.
While SN~2002ic has all the traits of a thermonuclear supernova it
also displayed a strong hydrogen Balmer H$\alpha$ emission line
\citep{ham03a}. This latter fact has led to an investigation whether
SN~2002ic could not be a core-collapse supernova \citep{ben06}.
SN~2006gy is a very energetic supernova clearly showing strong
H$\alpha$ emission, but a very slow light curve. One interpretation
argues for possibly the first observation of a pair-instability
supernova \citep{smi07}, while another study finds that this could be
a thermonuclear supernova within a dense circumstellar shell
\citep{ofe07}. Such cases show the difficulty to uniquely map the
classification scheme into the physical interpretation of the events. 

\section{Core-collapse supernovae}
\label{sec:typeii}

The richness in appearance of the core-collapse supernovae is due to
their varied progenitor histories. Spectra observed near maximum light
show H$\alpha$ in these events, but with a wide variety
\citep[e.g.][]{fil97,lei05}. The typical light curves of SNe~II
display a long plateau of about 100 days after the maximum. The most
prominent and best observed case after SN~1987A is SN~1999em
\citep{ham01a,leo01,elm03}. SN~1987A taught us a lot about core collapse
supernovae \citep[for reviews see][]{arn89,mcc93,lei03,mcc05}. While
SN~1987A displayed a strong P~Cygni line of H$\alpha$, it is almost
not visible in SN~1993J. The latter lost this line in its evolution
completely and only after about one year did H$\alpha$ reemerge in the
nebular spectrum \citep{fil94}.  The case of SN~1988Z is different
again. In this case, the hydrogen is excited in circumstellar material
shocked by the supernova ejecta. 
The emission is dominated by the shock energy and not recombination or
radioactive decay as in most other supernovae. 

This special case of supernovae interacting early on with their dense
circumstellar environment is discussed in \citet{che03} and
\citet{lei94}. They typically have very slow light curves and spectra
that show emission lines but very little absorption. The best studied
cases so far are SN~1986J \citep{lei91}, SN~1988Z \citep{tur93},
SN~1995N \citep{fra02} and SN~1998S \citep{fas00}. All of these
objects are strong radio emitters (reviews on the radio emission
are available from \citealt{wei88} and \citealt{wei02}). The radio
observations in particular allow to trace the mass-loss history of the
progenitor star with interesting conclusions on their final evolution.
These objects often can be observed for many years.
The poster child for a shock interacting with
circumstellar material is of course SN~1987A, which recently
transitioned from a teenager into a maturing supernova remnant
\citep{mcc93,mcc05,fra07}.

Extreme examples of this class of supernovae demonstrate their
diversity.  Examples are the GRBs (reviews in \citealt{wei03} and
\citealt{woo07a}), the
recent, very energetic SN~2006gy \citep{ofe07,smi07}, the very faint
objects like SN~1987A and the Type Ib/c events, which are presumably stripped
of their envelopes either by massive stellar winds or mass loss to a
companion star. All these different appearances are a signature of the
variety the evolution of massive stars leading to different 
configurations at the time of explosion. 

Several proposals have been made how core-collapse supernovae could be
used as distance indicators. They will be discussed in
\S\ref{sec:cosmo_ii}.

\section{Type Ia supernovae}
\label{sec:typeia}
Although thermonuclear supernovae have simpler underlying physics than
the core-collapse supernovae, there still remain formidable hurdles to
fully understand them \citep{hil00}. The observational material that has
been assembled in the last decade is considerable and many nearby
supernovae are now observed with exquisite detail. The last few years
have seen dramatic progress in recognising peculiar events and also
determining specific characteristics. The situation a few years ago is
described in \citet{lei00}. Since then the peculiar SN~2000cx
\citep{li01,can03} and SN~2002cx \citep{li03} have been observed. A truly
particular case has been discovered in SN~2002ic \citep{ham03a}, which
displayed a strong, broad-lined H$\alpha$ emission after about 90 days
past the maximum. This supernova displayed the signatures
of a bright SN~Ia with what looked like residual H$\alpha$ emission from
the host galaxy. The spectral sequence later showed that the hydrogen
emission is intrinsic to the supernova and indicates that this explosion
occurred inside a dense hydrogen cocoon. Such events throw a
dark shadow over the light curve {\it vs.} luminosity relations that have been
used in the past to normalise the peak luminosity
\citep{phi93,ham95,rie96a,rie98,per97,phi99,gol01,wan03a,wan06,guy05,guy07,pri06,
jha07} necessary to derive accurate cosmological
distances. The differences for individual objects highlight the fact
that not all SNe~Ia are identical and provide us with a tool to further
investigate the true nature of these explosions. 

The main observables of supernovae remain the optical and near-infrared
light curves and spectral evolution \citep[e.g.][]{lei00}.
Spectro-polarimetry in the optical has matured significantly over the
past decade and several SNe~Ia have significantly polarised light and
also remarkable evolutions \citep{kas03,wan03b,leo05,wan07}. Very few
observations at wavelengths outside the optical and near-infrared
window have been obtained. Only two events
have so far been observed in the thermal infrared, SN~2003hv and
SN~2005df \citep{ger07}. The detection of emission lines of nickel and
cobalt over 100 days after explosion indicates a surprisingly large
amount of stable nickel in the ejecta. Also, prominent lines of argon
([Ar~II] $\lambda6.985 \mu$m) with a double-horned profile are detected.
The observations hint at a stratified composition of the ejecta, which
cannot be explained well with the current models.
So far not a single SN~Ia has been
detected at radio wavelengths \citep{pan06} and only one X-ray
detection of a peculiar event has been reported (SN~2005ke,
\citealt{imm06}). 

Many optical and near-infrared light curves have become available.
Large collections of light curves are available
from the Cal\'an/Tololo and the Carnegie projects \citep[{\tt
http://csp1.lco.cl/$\sim$cspuser1/PUB/CSP.html}:
][]{ham95,phi99,phi06,phi07,kri01,kri03,kri04a,kri04b,kri04c,kri06,kri07},
the {\it CfA} group \citep[{\tt
http://www.cfa.harvard.edu/supernova/}:][]{rie96a,rie99a,jha06a}, the
Berkeley group \citep{fil92a,fil92b,li01,li03} and the more recent European
Supernova Consortium \citep[{\tt
http://www.mpa-garching.mpg.de/$\sim$rtn/}:][]{pig04,kot05,eli06,pas07a,pas07b,sta07,gar07a}. 

Most of the very early photometric observations have been provided by
these projects
(SN~2001el:
\citealt{kri03,kri07}; SN~2002bo: \citealt{ben04,kri04c}; SN~2003du:
\citealt{leo05,sta07}; SN~2004eo: \citealt{ham06,pas07a}; SN~2005cf:
\citealt{pas07b}) and the available data have more than doubled in the past
five years \citep{con06a}. The rise time appears to be roughly 18 days,
with some uncertainty whether there is a correlation with the light
curve decline rate as well \citep[e.g.][]{rie99b,con00}. 

Overall, the following picture has emerged for SN~Ia explosions. The
emission of SNe~Ia is powered by the stored energy in radio-active
decays from $^{56}$Ni through $^{56}$Co to $^{56}$Fe
(\citealt{col69,cla74}; see \citealt{kuc94} for an observational proof
of this mechanism for SNe~Ia). This release is moderated by the
optical depth in the ejecta \citep{arn82,hof93,pin00}.  Using Arnett's
rule \citep{arn82} one can derive the nickel mass from the observed
luminosity at peak light \citep{arn85,bra92a,vac96,con00,str05,str06}.
Not all Type Ia SNe produce the same amount of $^{56}$Ni in the explosions
\citep[e.g.][]{cap97,con00,str06}. Some objects are clearly
subluminous, a signature that very little radioactive nickel
is produced (most recent examples are SN~2002cx,
SN~2003gq, SN~2005P and SN~2005hk; \citealt{jha06b,phi07}). It has
been speculated that they are deflagration explosions rather than
delayed detonations. The derivation of the nickel mass based on
Arnett's rule has been tested from explosion models, hydrodynamics and
radiation transport calculations and has been shown to be reliable
\citep{bli06}. 

The interpretation of the light curves has seen a revival in the past
few years with attempts to explain the behaviour of the infrared light
curves, in which a second maximum is observed 
\citep[e.g.][]{eli85,mei00,kri03}.  The most convincing explanation is due to a
temperature sensitivity of the emissivity between singly and doubly
ionised iron-peak elements \citep{kas06a}. 
Depending on the temperature decrease in the ejecta, the energy is
released rapidly in the near-infrared and the secondary maximum is more
or less pronounced. A similar argument for a
temperature dependence in the SN~Ia spectra had been made by 
\citet{nug95} a decade earlier based on line ratios of Ca~II and Si~II. 

Further dependencies on the amount of nickel synthesised in the
explosion, the mixing within the ejecta and the progenitor metallicity
exist \citep{kas06a}. At the same time, these model calculations also
predict a very narrow distribution of the near-IR peak luminosity (based
on Chandrasekhar-mass models and a unique density structure of the ejecta),
as it is observed \citep{kri04a}.  There are now hopes that the light
curve width $vs.$ luminosity relation of SNe~Ia might be understood
through a detailed exploration of the parameter
space provided by current explosion models\citep{kas07}. 

At late times, the photometry and spectroscopy has been followed for
several objects. Especially the addition of the infrared has provided
new insights \citep{spy04,sol04,str07}. 
SNe~Ia have IR light curves, which after the peak phase are nearly
flat for several hundred days until the IR catastrophe sets in and the
ejecta cool enough so that the energy is radiated in fine-structure
lines in the thermal infrared rather than in the optical or the
near-infrared \citep{fra96}. As a consequence the IR contribution to
the bolometric flux increases dramatically 300 days after the
explosion.  Derivations based simply on the $V$ light curve (as
sometimes employed in the past) are hence unreliable at these late
phases. Also, the emerging flux is less than what is predicted
assuming Arnett's rule to determine the nickel mass from the peak
luminosity. This is a clear sign of $\gamma-$ray leakage from the ejecta
and a signature of low-mass progenitor stars. The late decline rate of
the light curves has been
used by \citet{str06} to crudely determine ejecta masses from the
bolometric light curves. The deviation of the decline rate from the
expected decay rate of $^{56}$Co is a signature of the losses due to
the decreasing column density in the ejecta. Using a very simple model
of the conversion of the $\gamma-$ray energy into the optical/IR
wavelengths the derived ejecta masses all are well below the canonical
Chandrasekhar-mass of the explosion models \citep{str06}. The reason
for this discrepancy remains unclear, but could be due to asymmetries,
i.e. dependencies on the viewing angle or a model that does not
capture the relevant physics. 

Another signature of variations in the explosions are
spectro-polarimetric measurements which show that certain elements in
the supernova ejecta are not distributed spherically \citep{wan03b,
leo05,cho06,cho07}. A synopsis of the current situation is given by
\citet{wan07}. It appears that there is only a small
asymmetry in the overall shape of the ejecta as the continuum
polarisation appears generally low and in most cases below the detection
limits. However, some stronger lines show a marked evolution in their
polarisation indicating that the material is not evenly distributed
throughout the ejecta and also giving clues on the possibly uneven
burning process. There even appears to be a correlation between the
degree of clumpiness and the luminosity of the supernovae with smaller
polarisations observed for more luminous supernovae \citep{wan07}.
 
The spectroscopic evolution has also obtained a lot of attention in the
past decade. Apart from some objects, which display truly different
spectra (in particular the cases of SNe~1999aa \citep{gar04}, 1999ac
\citep{gar05,phi06}, 2000cx \citep{li01}, 2002cx
\citep{li03,sol04,jha06b}, 2002ic \citep{ham03a,kot04}, and 2005hk
\citep{jha06b,phi07} should be mentioned here), the general spectral
evolution is characterised by different velocities at which the line
absorptions are observed. Detailed analyses of the velocity shifts go
back to \citet{bra88} and it is now established that most SNe~Ia show
high-velocity components in their spectra \citep{hat00,maz05}.
Observational trends appear to emerge in the way the velocities
within the supernova ejecta evolve \citep{ben05}, but the
interpretation of these correlations are not clear yet. It is noteworthy
that the distant objects appear to follow the general spectral
evolution
of their nearby counterparts and there is no obvious sign of differences
in the spectral appearance of SNe~Ia \citep{blo06,gar07b}. The
interpretation of the spectra has now also been expanded to reconstruct
the element distribution in the ejecta through the spectral evolution
\citep{fis99,ste05}, which gives a direct input to the explosion
models. Also, spectral calculations based on non-spherical ejecta are
leading to new explanations for the luminosity and expansion velocity
variations in SNe~Ia \citep{kas06b,sim07a,sim07b}.

The ideas on the explosion models have evolved only little in the past
few years. The favourite mechanisms are the delayed detonation, in which
an early deflagration (burning slower than the local sound speed) turns
into a detonation (burning front moves supersonically) in the out layers
\citep{kho91,rop07a,rop07b}, and pure deflagrations (see \citealt{hil00} 
for a review of these models). Deflagrations in general are regarded as
not providing enough energy for the brilliant displays of SNe~Ia,
however, in a few cases would a simple deflagration provide sufficient
energy for a SN~Ia \citep{bli06,jha06b,phi07}. There has been a lot of
activity in extending the calculations into full three-dimensional
simulations to explore the effects of asymmetries
\citep{rei02,gam03,gam04,gam05,rop05,rop06}.  The simulations are
now also incorporating off-centre ignitions and other aspects, which
could lead to non-uniform explosions \citep{sim07b}. 

Despite these advances, it remains to be understood, why SNe~Ia can be
calibrated with rather simple methods to provide accurate cosmological
distances. 

\section{Cosmology with Supernovae}
\label{sec:cosmo}
Cosmology with supernovae has developed over the second half of the
last century. 
Various methods were devised to use supernovae to determine
cosmological parameters ranging from simple standard candle paradigms
to physical explanations of the supernova explosions and subsequent
derivation of distances. The simplest use has been the determination
of luminosity distances, i.e. the comparison of the observed flux to
the total emitted radiation. A more elaborate method is the comparison
of the angular diameter, through the measurement of the radial
velocity of the expanding atmosphere, and the observed brightness. A critical
assumption here is the sphericity of the explosion and the
corresponding connection of the ejecta velocity and the luminosity,
which has to be achieved through detailed emission models of the
supernova explosion. 

The classical parameters of observational cosmology, which govern the
expansion of the universe in Friedmann-Robertson-Walker models, the
Hubble constant H$_0$ and the deceleration parameter q$_0$, can be
determined with accurate (luminosity) distances
\citep{san61,san88,wei72,pee93,pea99}. There is a rich literature on
the Hubble constant and Type Ia supernovae 
(see \citealt{bra92b,bra98,lei01a,per03} for reviews). The deceleration
parameter has been replaced by more modern formulations specifically
including the cosmological constant or some variants thereof
\citep{car92} and is generally referred to as 'Dark Energy.' Detailed
theoretical descriptions are given in other articles of this issue. 

\subsection{The Hubble constant}
\label{sec:H0}

\subsubsection{Core-collapse supernovae}
\label{sec:cosmo_ii}

Following early work by \citet{baa26}, originally done for Cepheid
stars, the expanding photosphere method
\citep[EPM;][]{kir74,sch94,eas96,ham01b,ham02,des05} has been applied
to several supernovae. The most comprehensive data sample has been
assembled by \citet{ham01a}. A critical test has become the distance
to SN~1999em, which was determined through EPM
\citep{leo01,ham02a,elm03,baro04,des06} and which also has a Cepheid
distance available \citep{leo03}. The discrepancy in the distance
determinations towards SN~1999em can be attributed to the fact that the
correction factor for the dilution of the black body flux in EPM are
strongly model dependent and need to be calculated for each supernova
individually \citep{baro04,des05}.

Recently, Mario Hamuy has realised that the expansion velocity and the
luminosity during the plateau phase correlate and that Type II SNe may
be calibrated to become quite good distance indicators \citep{ham02}.
The distance accuracy achieved this way can be better than 20\%. These
determinations are based on the physical understanding of the plateau
phase of SNe~II and are linked to physics of the supernova atmosphere.
This means that they are independent of the {\it distance ladder},
which is the basis for the SNe~Ia (see \S\ref{sec:cosmo_ia}). Typical
values for the Hubble constant from SNe~II are in the range of 65 to
75 km~s$^{-1}$~Mpc$^{-1}$ \citep{ham03b}.

A first attempt to derive the Hubble diagram with distant (up to
z$\sim$0.3) SNe~II using data assembled by the {\it CFHT} SN Legacy Survey
has also been made recently \citep{nug06}. Potentially, this method can
independently check on the cosmic expansion history.

\subsubsection{Type Ia supernovae}
\label{sec:cosmo_ia}

The best way to show that objects provide good relative luminosity
distances is to plot them in a Hubble diagram. Originally, this
diagram was using recession velocity {\it vs.} apparent magnitude
\citep{hub36,san61}. The underlying assumptions are that the Hubble
law holds, i.e. the local expansion is linear, and that the objects
are all of the same luminosity, i.e. standard candles, so that the
apparent brightness directly reflects distance. Early
versions of this Hubble diagram of SNe~Ia showed that the peak
magnitudes tracked the Hubble line fairly well
\citep{kow68,tam90,lei92}, but considerable scatter was still present.

There are essentially three quantities that can be derived from such a
Hubble diagram in the nearby universe: the slope of the expansion line,
the scatter around the expansion line and the value of the local Hubble
constant from the intercept at zero redshift 
\citep[e.g][]{tam90,lei92,bra92b,rie96a,bra98}. The slope gives an indication of
the local expansion field and for a linear expansion in an isotropic
universe has a fixed value. The scatter around the expansion line
provides a measure of the accuracy of the relative, in contrast to an
absolute, distance determination, individual deviations from the smooth
cosmological expansion and the measurement errors. The intercept of the
line, finally, together with an estimate of the absolute (normalised)
luminosity provides absolute distances and hence the Hubble constant.
Recent Hubble diagrams of SNe~Ia have been published by
\citet{ton03,kno03,barr04,rie04a,rie04b,ast06,woo07b,rie07} and
\citet{jha07}. It
should be noted that SNe~Ia may be nearly standard candles in the
near-infrared \citep{kri04a}. The first significant IR sample shows very
small scatter without prior correction for light curve shape. 

Modern versions of this diagram have exchanged the recession velocity
with the redshift, often corrected to the CMB rest frame and the
distance modulus instead of the simple observed apparent peak
brightness. It has become clear that SNe~Ia are not simple standard
candles (see \S\ref{sec:typeia}, an extensive discussion is given in
\citealt{lei04}). Hence, the distance has to be determined for each event
individually, e.g. through the maximum luminosity {\it vs.} light curve
width relation discussed in \S\ref{sec:typeia}. Another option is to
normalise the peak luminosities and to plot a 'corrected' apparent peak
brightness, a method employed by the Supernova Cosmology Project 
\citep[e.g.][]{per97,per99,kno03}. This approach is masking the importance of
the light curve correction and also the importance of the absorption
corrections. 

\begin{figure}[htb]
\centering
\includegraphics[width=8.4cm, angle=270]{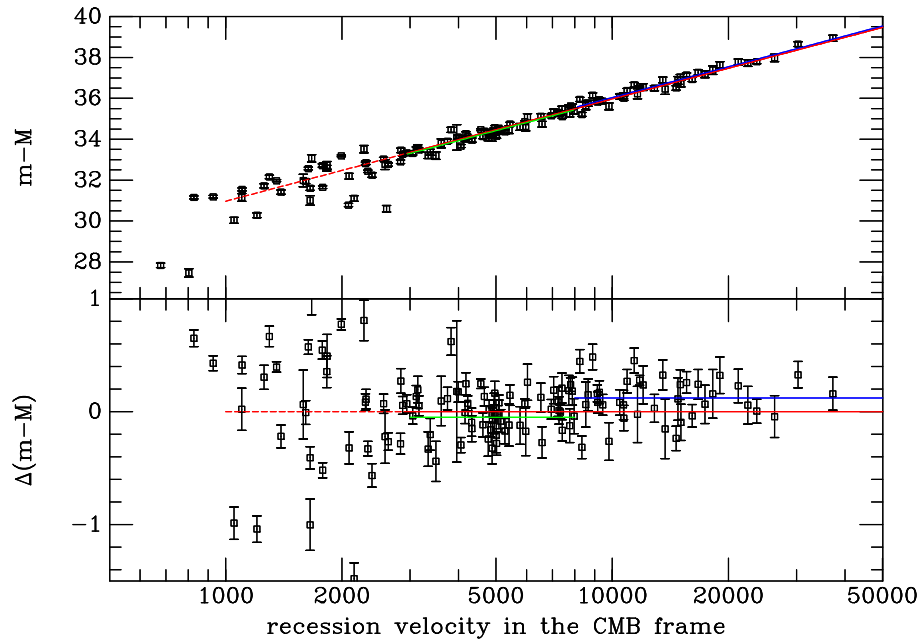}
\caption{Hubble diagram of nearby Type Ia supernovae. The distances are
derived from light curve shape corrected luminosities 
\citep[data from][]{jha07}. Fits to different velocity ranges are
shown. The red line is a fit to all SNe~Ia with v$>$3000~km~s$^{-1}$
(extrapolated to lower velocities as a dashed line),
the green line for the sample restricted to
3000~km~s$^{-1}$$<$v$<$8000~km~s$^{-1}$ and the blue line for events
with v$>$8000~km~s$^{-1}$.}
\label{fig:hub_near}
\end{figure}

The scatter of the normalised SNe~Ia around the linear expansion line
is less than 0.2 magnitudes or 10\% in distance
(\citealt{phi99,jha99,ton03,rie04b,jha07}; Fig.~\ref{fig:hub_near}).
Independent of our ignorance of the exact explosion mechanism or the
radiation transport in the explosions this proves that SNe~Ia can
reliably be used as a (relative) distance indicator in the local
universe and makes them empirically calibrated.  This situation is
very much comparable to the Cepheid stars, where the period-luminosity
relation is based on empirical data from objects in the Magellanic
Clouds.

Fig.~\ref{fig:hub_near} displays the most recent, homogeneously treated
sample of nearby SNe~Ia from \citet{jha07}. The upper panel displays the
regular Hubble diagram with distance {\it vs.} recession velocity
corrected to the rest frame of the cosmic microwave background, while
the lower panel shows the data with the expansion field removed. This
allows to appreciate the accuracy of the relative distances
derived by the supernovae and also
provides a better demonstration of the various cosmological models. This
will become even clearer for the full Hubble diagram discussed below
(Fig.~\ref{fig:hub}). The distance modulus ($m-M)$ combines the observed
magnitude with the observed flux $F$ through $$m=-2.5\log(F)+const$$ and
the absolute luminosity $L$ of an object at the distance of 10pc
$$M=-2.5\log(L)+const$$ and is determined for each supernova
individually. The distance modulus describes the observed flux ratio
of two objects at different distances
according to the usual 1/D$^2$ law, which defines the cosmological
luminosity distance and the observed flux with distance and emitted
energy through $$F=\frac{L}{4\pi D^2}.$$ 

For a linear cosmic expansion following Hubble's law $$D=\frac{v}{H_0}$$
one expects that the distance moduli and the recession velocities are
connected through  $$(m-M)= 5\log(v)-5\log(H_0)+25$$ where the velocity
is measured in km~s$^{-1}$, the distance in Mpc and the Hubble constant
H$_0$ has units of km~s$^{-1}$~Mpc$^{-1}$.

It is obvious in Fig.~\ref{fig:hub_near} that below a recession
velocity of about 3000~km~s$^{-1}$ the supernovae do not trace the
smooth Hubble expansion, but the Hubble flow is heavily disturbed by
motions due to the local matter distribution, often referred to as
'peculiar velocities.' These supernovae are regularly excluded from the
cosmological studies. The slope above 3000~km~s$^{-1}$
is slightly larger, i.e. $5.22\pm 0.05$, than the expected value for
the linear expansion in the local universe, which could be an
indication of evolution. 

We demonstrate in Fig.~\ref{fig:hub_near} the effect of a possible
change in the universal expansion rate at some distance from us. The
lower panel shows the fits to data with $v>3000$~km~s$^{-1}$, a fit to
the data in the range $3000<v<8000$~km~s$^{-1}$ and the data with
$v>8000$~km~s$^{-1}$, where we force the fit for a linear expansion.
The upper value was taken to be close to the reported outer edge of a
possible 'Hubble bubble' \citep{zhe98,jha07} where the expansion
inside is faster than outside and hence the true Hubble constant would
be lower than what is determined locally. Indeed, there appears to be
a shift by about 0.07 magnitudes (about 4\% change in H$_0$) for the
objects outside 8000~km~s$^{-1}$. Another interpretation traces this
change to an evolution in the intrinsic colours of SNe~Ia
\citep{con07}. 

By fitting the intercept of the expansion line a combination of the
Hubble constant and the absolute luminosity is determined. Hence, for
the derivation of the Hubble constant the (normalised) luminosity of the
SNe~Ia has to be known. The most direct way to achieve this is through
the distance ladder and in particular the calibration of nearby SNe~Ia
by Cepheids \citep[for the most recent
results see][]{sah99,fre01,san06}.
The main discrepancy for the published values of the Hubble constant
from SNe~Ia is coming from the different interpretations of the Cepheids
and application of the light curve shape correction. Ironically, the
SNe~Ia provide the best distance indicator beyond the Cepheid range and
have replaced many rungs in the distance ladder making the Magellanic
Clouds the last rung before cosmological distances. We do not quote a
value for the Hubble constant here. The interested reader is referred to
the papers mentioned above.

A different way to establish the Hubble constant with SNe~Ia is through
models. Originally tried by \citet{arn82} and \citet{arn85} this has
been further attempted by \citet{lei92} and most recently by
\citet{str05}. In this case the absolute luminosity is derived from the
amount of nickel produced in the explosion models and the derived luminosity,
e.g. through Arnett's rule or direct radiation hydrodynamics
calculations. Due to the range of observed SN~Ia properties it is not
possible to derive a value for the Hubble constant itself, but at least
an interesting lower limit of $H_0>50$~km~s$^{-1}$~(3$\sigma$) could be
derived by matching the faintest observed SNe with the largest
imaginable nickel mass ($\sim$1~M$_{\odot}$) for the models. Overall, a
slight inconsistency between the predications of the current models and
the observations could be found. By adopting a Hubble constant of
$\sim$70~km~s$^{-1}$~Mpc$^{-1}$ one can derive a predicted range of
nickel masses in the explosions ($0.5 {\rm M}_\odot < {\rm M}_{\rm Ni}<
1.0 {\rm M}_\odot$; \citealt{str05}).

\subsection{The expansion history of the universe}
\label{sec:expansion}
Exploring the cosmic expansion rate over the history of the universe
tells us about the changing contributions of the different matter/energy
components of the universe (see the article by Linder). The
supernovae provide an important information by mapping out
the expansion history over a significant lookback time (out to a
redshift of z$\sim$1.5, corresponding to a lookback time of about 2/3 of
the age of the universe, or over 9~billion years for 
the concordance model and $H_0=70$~km~s$^{-1}$~Mpc$^{-1}$). It
should be stressed that for the expansion history only relative distances
are need to be measured. The SN~Ia Hubble diagram of nearby objects
(Fig.~\ref{fig:hub_near}) gives ample empirical confidence that
this can be achieved reliably. 

The published distances of high-z supernovae are typically based on an
{\it adopted} Hubble constant. Several theoretical papers in the recent
past have made the mistake to include the Hubble constant as a free
parameter in their fits. While this is okay to check that the
marginalisation actually works correctly, claims that a specific
value for the Hubble constant has been found are incorrect. The original
papers all state very clearly what Hubble constant has been adopted for
the study and people who use those data should be aware of this
assumption. 

The proposal to use supernovae to measure the cosmic deceleration goes
back to Olin \citet{wil39} and was elaborated further by
\citet{tam78} and \citet{col79}. one prediction
made by these early visionaries was that time dilation would affect
the observed light curves. This could finally be shown convincingly
with the first distant SN~Ia, SN 1995K, by \citet{lei96} and was
further confirmed on a large sample by \citet{gol97,gol01}. In the
meantime this test has been performed following the detailed spectral 
evolution \citep{rie97,fol05,hoo05,blo06,blo07}. The
predictions of a universal expansion have been confirmed in all cases
ruling out alternative theories of ``tired light.''

Proposals to use SNe~Ia to measure the expansion history of the universe
go back into the late 1980s. The main goal at the time was to
determine the mean matter density $\Omega_M$ to check the cosmological
models. The first observational attempts were frustrated by lack of
'grasp,' i.e. the difficulty to cover large enough area on the sky to
sufficient depths frequently enough. A search with the Danish 1.5m
telescope on La Silla monitoring several fields once per month yielded
only two distant SNe after two years. The follow-up
spectroscopy was difficult to organise in a time before observatories
were fully connected to the Internet and the information had to be
transmitted through fax and telex, a particular problem for finding
charts. The spectroscopic capabilities of the
available 4m telescopes were marginal for the faintness of the
objects \citep{nor89,han89,sch98,rie98}. A large project to search for
distant SNe~Ia was initiated in the early 1990s in Berkeley
\citep{per91} and yielded first results on seven objects (several
without spectroscopy and insufficient colour coverage \citealt{per95}). As
a result the inferred cosmology was not correct \citep{per97}. The
following years saw the emergence of vastly improved search techniques,
the advent of 8m and 10m telescopes --- greatly improving the quality of
the spectroscopic confirmations, refined analysis methods taking
many contaminating effects into account and the delivery of a surprise.
With the proof of concept from the early searches the new
projects, the Supernova Cosmology Project
\citep{per95,per97,per98,per99,kno03,hoo05} and the High-z Supernova
Search Team
\citep{sch98,lei96,rie97,rie98,gar98a,gar98b,rie00b,coi00,ton03,wil03,barr04,clo06},
started to provide astonishing evidence that the distant SNe~Ia appeared
fainter than predicted in a massless, empty universe. Early criticism of
these results concentrated on difficulties with photometric accuracy of
the faint sources, the treatment of the dust absorption in the host
galaxy of the supernova, possible secular evolution of the supernovae
over time, uncertainties in the normalisation of the peak luminosity
of the SNe~Ia and the, at the time still fairly small, sample size of
distant objects, which could lead to sample biases or Malmquist effects
(see \citealt{lei01a} for a summary of these early problems). Exotic
possibilities, like unusual dust properties \citep{agu99a,agu99b} were
proposed or difficulties with the normalisation pointed out
\citep{dre00,lei00}. Many of these difficulties have been addressed in
the meantime.  Also, the importance of the nearby SN~Ia sample should
not be underestimated. The reason that \citet{rie98} could find a signal
for accelerated expansion with only 10 distant SNe~Ia was largely due to
the fact that an extensive, controlled, local sample of SNe~Ia was at
hand. 

In the past few years the Canada-France-Hawaii Telescope {\it CFHT}
Supernova Legacy Survey (SNLS; {\tt http://www.cfht.hawaii.edu/SNLS/})
and the ESSENCE project ({\tt http://www.ctio.noao.edu/wproject/})
have been collecting data of distant supernovae to measure the value
of a constant equation of state parameter $\omega$ to 7\% and 10\%
accuracy, respectively. The SNLS monitors four fields with the MegaCam
instrument at the {\it CFHT} continuously, while ESSENCE uses the MosaicII
camera with the CTIO 4m telescope during three months each year. The
ultimate goals of these five-year projects are $>$700 SNe~Ia for SNLS
and $>$200 SNe~Ia for ESSENCE. All supernovae must have a positive
spectral classification to be included. 

The SNLS has published cosmological results of their first
year of observations based on 71 distant SNe~Ia \citep{ast06}. The
selection of the candidates and the spectroscopy of this project are
described in \citet{sul06a,lid05} and \citet{how05}. Other important
results based on this extensive data set are a determination of the
SN~Ia rise time \citep{con06a}, as well as the supernova rates and their
connection to star formation in the host galaxy \citep{sul06b,nei06}.
Further, this project obtained observations of a peculiar SN~Ia possibly
emerging from a super-Chandrasekhar-mass progenitor \citep{how06} and
made a first measurement of distances at $z>0.1$ of SNe~II
\citep{nug06}. 

The ESSENCE project is presented in \citet{mik07} and the cosmological
results based on the first three years including 60 SNe~Ia are
discussed in \citet{woo07b}. All corresponding spectroscopy has been
published \citep{mat05,blo06,blo07}. A first detailed description of
photometry of a subset of the ESSENCE events observed with the Hubble
Space Telescope ({\it HST})
pointed out some potential selection effects in the sample
\citep{kri05}. An evaluation of exotic proposals for dark energy when
compared to the available SN~Ia data was made in \citet{dav07}. 

A separate project including many ESSENCE members is the higher-z SN
search with {\it HST}. The targets for this study have been SNe with z$>$1
\citep{str04}. These high-z supernovae have shown that the universe
indeed was decelerating at z$>$1 and the acceleration phase has started
only during the second half of the universal history
\citep{rie04a,rie04b,rie07}. The most recent data sample allowed
\citet{rie07} to map out the change of the Hubble parameter over
redshifts for the first time ever directly showing that the universal
expansion rate has changed over time. This project also yielded
important results on the evolution of the SN~Ia rate as a function of
redshift \citep{dah04}. However, the inference of long lead time
before a SN~Ia explosion has been disputed \citep{for06}.

Other ongoing projects are the continuation of the Supernova Cosmology
Project ({\tt http://panisse.lbl.gov/ACSclustersearch/}) to find
supernovae in distant clusters with z$>$1. The goal is to observe SNe~Ia in
elliptical galaxies as the problem with the extinction in the host
galaxy is strongly reduced. A first exploration of this method had been
done by \citet{sul03}. The claim has been made that SNe~Ia in
elliptical galaxies provide a cleaner sample. Possible problems with this
approach is the lack of a good comparison sample of local supernovae.
Data for a first object have recently been published exploring new
ground-based observational methods, in particular adaptive optics
imaging \citep{mel07}.

The extension of the Sloan Digital Sky Survey for a three-year supernova
search is ongoing ({\tt
http://sdssdp47.fnal.gov/sdsssn/sdsssn.html}). The goal is to find 200
SNe~Ia at 0.1$<$z$<$0.3. This project appears to be quite successful with
many spectroscopically confirmed SNe~Ia. An impressive mosaic is
available from the above Web page and has been published in National
Geographic Magazine. The local supernova searches have been described in
\S\ref{sec:typeia}. One should add here the SN Factory ({\tt
http://snfactory.lbl.gov/}), which is
specifically set up to provide a large sample of nearby SNe~Ia for the
comparison with the high-z sample. So far only few events from this
project has been published \citep{ald06,tho07}, all of peculiar nature. 

The cosmological signal imprinted on the supernova data is modulated
by several unwanted technical and astrophysical effects.  At the basis
is accurate photometry \citep{stu06}. While this sounds like a trivial
statement, it has become difficult to free the measurement from all
the effects of Earth's atmosphere to the percent level required for
the SN light curves. Improvements in the instrumental
characterisations are made continuously \citep{mik07}, but one of the
limiting effects are the implementation of the various filter
pass bands at the telescope, which has to be known accurately to be
able to combine observations from different telescopes \citep{dav06}.
For nearby supernovae this led to the introduction of an empirical
correction (often referred to as S-correction) of data sets from
different telescopes \citep{str02}.  As a consequence recent projects
concentrate on single instruments ({\it CFHT}/MegaCam for the SNLS and CTIO
Blanco telescope/MosaicII for ESSENCE) for the photometry to avoid
this problem.  Nevertheless, it still remains difficult to combine
SNLS and ESSENCE data for a joint analysis as done in \citet{woo07b},
\citet{rie07}, \citet{dav07} and various other publications. 

Since the supernovae have to be corrected for foreground extinction
the colour needs to be measured as accurately as possible. Any
uncertainty in this respect is multiplied by the absorption
correction. The uncertainty of the colour measurement also has a
direct influence on the K-correction \citep{ham93,kim96,nug02,hsi07}. The
observed photometry has to be translated into the supernova rest frame
and hence any redshift of the spectrum needs to be taken into account
(see \citealt{jha07} for a detailed description of this problem and a
current implementation). The K-corrections are time-dependent and need
to be calculated for the correct phase as well as the correct
intrinsic colour. This intimately connects the K-corrections with the
absorption correction and modern versions of light-curve fitting
programmes for distant supernovae merge this evaluation. The light
curve fitting methods and calibration are critical to the supernovae
cosmology and it should be emphasised that depending on which methods
are used, the derived distances can change -- sometimes in a
systematic way.  \citet{woo07b} have performed a detailed analysis of
the SNLS/SALT fitter \citep{guy05} and MLCS2k2
\citep{rie04b,rie07,jha07} and confirmed that consistent cosmological
results are derived by the two methods on the same data sets, but
small differences remain. 

Detailed spectroscopy certainly would help, but the signal achieved with
the current telescopes is still limited and in many cases the supernova
spectrum is contaminated by host galaxy light. Methods to separate the
SN spectrum from the galaxy are either to try a deconvolution
\citep{blo05} or subtract a scaled galaxy spectrum \citep{sai04,how05}.
The spectroscopy is also essential to distinguish SNe~Ia from luminous
SNe~Ib/c as the two classes stem from distinct explosion mechanisms and
confusion could lead to wrong conclusions, when the objects are not
separated correctly \citep{hom05,tau06}. 

Astrophysical effects, which can influence the cosmological
interpretation of supernova data include absorption in the Milky Way and
in the host galaxy, gravitational lensing, evolution of the supernovae
as a function of age of the universe, e.g. due to different
metallicity, selection biases due to limiting
sampling of the intrinsic supernova distribution and effects from a
local underdensity, which would mean that the local expansion rate is
lower than the global one ('Hubble bubble'). 

Several of these are well under control. Gravitational lensing does not
appear to be a major issue for the redshifts considered so far. The
highest redshift supernovae may be affected by lensing individually, but
the overall effect should be minimal
\citep{wam97,hol98,ama03,gun06,jon06,jon07}. The absorption due to dust
in our own Milky Way is also fairly easily corrected. The effect is
somewhat
alleviated by the redshift and the diminished influence of dust
absorption at redder wavelengths. Evolution of the supernova peak
luminosity could mimic a cosmological effect, but the available data
do not indicate any significant changes between the local and distant
SN~Ia samples. Within the achievable accuracy
the distant supernovae appear the same spectroscopically
\citep{hoo05,lid05,mat05,blo06,rie07} and also their light curve
behaviour appears rather similar to the local sample
\citep{ast06,woo07b}. The effect of the metallicity of the progenitor
star is predicted to be insignificant \citep{rop04}.

Our deductions on cosmology and dark energy could be severely hampered
by the limited accuracy with which we know the local expansion field
\cite[see \S\ref{sec:cosmo_ia},][]{hui06,coo06,jha07}, selection biases
which skew the observed distribution from the intrinsic one
\citep{lei01a,woo07b}, and our
lack of a good understanding of the dust properties in the host
galaxies \citep{eli06,ast06,woo07b}. 

As already shown in \S\ref{sec:cosmo_ia} the local expansion field is
not smooth and local flows are distorting our ability to set the
zero-point for the expansion rate. This leads to a systematic
uncertainty, which needs to be overcome, if more accurate
determination of cosmological parameters will be attempted. The effect
is of the order of about 6 to 8\% overall \citep{jha07,woo07b}.
Larger nearby supernova samples are required to evaluate the reality
of a Hubble bubble. Another possibility is to improve our knowledge
of the density distribution in the local universe (as attempted over
a decade ago, e.g. \citealt{ber90,bla99}) for a better understanding
of the local disruption of the smooth universal expansion. With very
large samples of nearby supernovae one could attempt to map this
density field as well, but that will likely require several thousand
supernovae. 

The problem can of course also be inverted and the SNe~Ia be used to
determine the local velocity field compared to the CMB. This has been
done with early samples by \citet{rie95} and more recently by
\citet{hau07} who find a quadrupole in the velocity distribution. 

Ideally one would like to use distance limited supernova samples. With
a hypothetical standard candle, which has a narrow luminosity
function, one would hope that a flux limited sample would also be
volume limited.  There are several reasons why the available distant
supernova samples are not volume limited. First, the SNe~Ia are not
standard candles and their luminosity function is spanning almost a
factor of 10 from the brightest to the faintest events. Even though
the most extreme cases are not included the most distant supernovae
are also the most luminous ones \citep{kri05}. Second, supernova
searches all use a certain frequency, with which the search fields are
monitored. This means that a supernova is discovered during its rise
and depending on the distance and the weather conditions objects will
be lost \citep{mik07}. Finally, dust absorption in the host galaxy
will dim some events, hence make them too faint to be discovered and
remove them from the sample. A priori this would seem not such a
problem, but it turns out that for more distant objects this becomes
progressively more important and together with the limited sampling
frequency creates a systematic bias. \citet{woo07b} have simulated
this effect in detail for the ESSENCE data set and found a
considerable bias (nearly 0.3 magnitudes in distance modulus at
z=0.6), if the default absorption prior was used. They introduced 
separate, redshift dependent priors for the ESSENE data to correct for
the fact that more SNe~Ia go undetected at higher redshift and larger
host galaxy absorption. The classical Malmquist bias is here mixed
together with the assumption on the intrinsic colours of the
supernovae and the absorption in their host galaxies. 

The unknown reddening law in external galaxies is a further
uncertainty, which systematically limits our ability to determine
cosmological parameters. Light scattering depends on the physical size
of the dust particles. So far the local absorption law has been
assumed for all supernovae, but it has been shown that for many
heavily extincted SNe~Ia a different reddening law seems to apply
\citep{rie96b,kri00,eli06,ast06}. The curious fact is that with the
regular colour dependence a colour excess, i.e. an apparent redder
colour due to the interstellar dust scattering, is rather large for
bluer bands. The canonical value for the solar neighbourhood is about
3.1 for the visual {\it V} band. For many SNe~Ia this value appears to
be reduced to somewhere between 2 and 3. This has also the curious
effect that the absorption correction for some supernovae is reduced
and the scatter in the distances reduced. However, once the reddening
law is a free parameter it can be assumed that it will be vary for
different sight lines through distant galaxies. This will introduce a
random scatter, which will be very difficult to overcome. For the
ESSENCE supernovae, these combined colour effects constitute the
largest uncertainty (about 10\% overall; \citealt{woo07b}). 

These last uncertainties will not easily be remedied by larger
samples. They present fundamental shortcomings of our understanding of
some of the critical items in supernova cosmology. They are not
directly related to the supernova physics itself, but are an
expression of the fact that the universe is filled with clumped matter, dark
and baryonic, which distorts our position as a fair observer of the
universe and affects the light we observe of these distant objects.
Overcoming these systematic difficulties will be key to further improve the
accuracy with which we can determine the cosmological parameters. 

\begin{figure}[htb]
\includegraphics[width=8.4cm, angle=270]{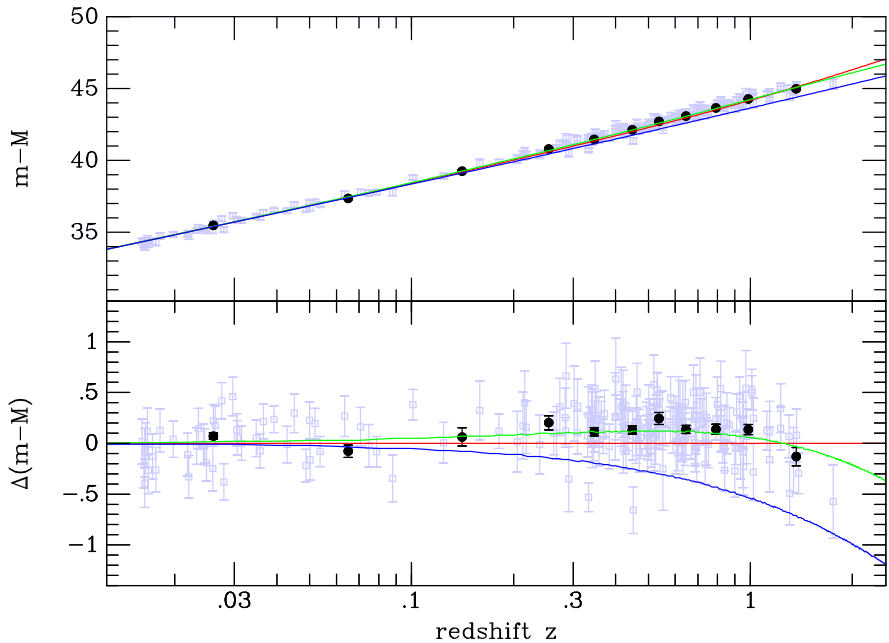}
\caption{Hubble diagram of Type Ia supernovae. The distances are
derived from light curve shape corrected luminosities 
\cite[data from][]{dav07}. The red line is for an empty universe
($\Omega_{\Lambda}=\Omega_M=0$), the blue line for an Einstein-de
Sitter model ($\Omega_{\Lambda}=0, \Omega_M=1$).
The concordance model
($\Omega_{\Lambda}=0.7, \Omega_M=0.3$) is shown as the green line fitting the
data best. The bottom panel shows all distances relative to the empty
universe model. The data for the individual supernovae is plotted as
shaded point, while the binned data are shown in black.  }
\label{fig:hub}
\end{figure}

Figure~\ref{fig:hub} displays the latest data set, which is a
combination of the largest nearby SN~Ia sample from \citet{jha07}, the
ESSENCE data \citep{woo07b} and the published SNLS \citep{ast06}. The
data are remarkably consistent with the concordance model of
$\Omega_M=0.3$ and $\Omega_{\Lambda}=0.7$. 

The SNe~Ia are further used to determine the integral of the equation
of state parameter $\omega$ over the observed redshift range
(z$<$1.7). All experiments find a consistent value of $\omega=-1$
within the uncertainties. Currently these are of about 13\%
statistical and 13\% systematic for ESSENCE \citep{woo07b} and 9\%
statistical and 5\% systematic for SNLS \citep{ast06}. These values
are unfortunately not directly comparable as different assumptions
went into the calculations of the errors. Nevertheless, all results so
far are consistent (within 1 $\sigma$) with a cosmological constant.
An important ingredient in this derivation is the matter density,
which in most recent studies has been taken from the baryonic
acoustic oscillation measurements of \citet{eis05} or \citet{col05}.
The accuracy of the derivation of $\omega$ strongly depends on how
well the matter density $\Omega_M$ can be constrained. Sometimes a
flat geometry of the universe is also assumed. 

Attempts have been made to derive constraints on a possible time
dependence of $\omega$ using the supernova data. One should caution
these enterprises as they are based on data, which are most likely not
accurate enough to warrant such analyses. Most published attempts
demonstrate this fairly clearly as the parameters become essentially
unconstrained \citep{rie04b,woo07b,rie07}. Several theoretical papers
have further ignored the systematic uncertainties in the data and may
have derived spurious results. 

One other interesting application of the Hubble diagram of SNe~Ia is the
attempt to constrain any change of Newton's gravitational constant G.
The current limits exclude changes larger than
$|\frac{\dot{G}}{G}|<2.9\cdot10^{-11}$~year$^{-1}$ \citep{gaz02,lor03,gar07}.

\section{Outlook and future projects}
\label{sec:fut}

SNe~Ia are amongst the most promising candidates to further improve
our view of the cosmos. They appear prominently in the recent studies
on how dark energy could be further constrained \citep{alb06,pea06}. 
Together with other probes of the deep universe the SNe~Ia should help
us to characterise dark energy and possibly discover its nature. 

Several projects have been proposed. The next surveys require new
instrumentation, in particular wide-field cameras and dedicated
telescopes. The SNLS has already shown the way forward with its
allocation of several hundred nights on a single telescope. The next
step is the Dark Energy Survey ({\tt
http://www.darkenergysurvey.org/}) planned with the CTIO Blanco 4m
telescope. For this project a new camera is being built for this
telescope. The goal is to observe 2000 SNe~Ia with 0.3$<$z$<$0.8.
Future survey telescopes like the Large
Synoptic Survey Telescope ({\it LSST}; {\tt http://www.lsst.org}) or
The Panoramic Survey Telescope \& Rapid Response System ({\it
Pan-STARRS}; {\tt http://pan-starrs.ifa.hawaii.edu/public}) will find
thousands of supernovae. It will become impractical to obtain
spectroscopy for all these objects for the classification and
statistical approaches using the observed light curve shapes and the
colours are being developed
\citep[e.g.][]{barr04a,rie04b,sul06a,con06b,kuz07}. However, it still
needs to be demonstrated that such large samples will allow us to
improve the cosmological parameters. 

An important extension of the current supernova work is towards higher
redshifts. The sample of known SNe~Ia at z$>$1 is very small still
\citep{rie07} and these events help significantly to constrain the
cosmological models and also to check for systematic effects in the
supernovae. All these very distant SNe~Ia have been found by {\it HST}
and its large area Advanced Camera for Surveys {\it ACS}. This is one
reason why future space projects aim at wide field imaging. The
synergy with weak lensing studies are obvious and strong science
drivers for these missions have been developed. The best know proposal
is the SuperNova Acceleration Probe (SNAP; {\tt
http://snap.lnbl.gov}), which has stimulated many interesting studies
of what could be achieved by such a data set.  Currently the SNAP
satellite could reach SNe~Ia out to z$\approx$1.5. Three missions have
been selected for a study as a Joint Dark Energy Mission (JDEM)
between NASA and the US Department of Energy. They are the Advanced
Dark Energy Physics Telescope (ADEPT), the Dark Energy Space Telescope
(Destiny; {\tt http://destiny.asu.edu}) and SNAP. All of them employ
the supernova Hubble diagram in addition to weak lensing surveys to
further characterise dark energy. 

To overcome the difficulties with the optical colours it has been
suggested to construct a supernova Hubble diagram in the near
infrared. At these wavelengths the SNe~Ia are showing very small
scatter in their peak luminosity and promise to approach the standard
candle concept better than at the blue wavelengths employed so far
\citep{kri04a}. The difficulty so far has been that due to the
redshift the rest frame near-infrared wavelengths are pushed to
wavelengths were not enough sensitivity is available. With the future
{\it JWST} and its infrared capabilities it will be possible to
compile a Hubble diagram of distant SNe~Ia in the near infrared. This
will present a critical test of the current results and may
significantly improve the distance accuracy as several limiting
effects, like light curve shape and reddening corrections can be
avoided. A first attempt of a Hubble
diagram in the {\it I} pass band has been made by \citet{nob05}. 

An independent test of the cosmology will come from an extended Hubble
diagram of type II supernovae. These distances are based on completely
different physical assumptions. Work in this direction has started
\citep{nug06}. 

Further improvements will come from a better understanding of the
explosions themselves. The question whether the distant SNe~Ia are
identical to the ones observed locally has not been fully addressed.
The currently available observational resources do not allow us to
obtain data of the required quality to compare, e.g., the spectral
evolution of the distant supernovae. With a secured model for the
explosion, it will become easier to explore possible systematic
differences of supernovae coming from younger progenitor systems than
older ones. There have been discussions of differences between
supernovae coming from presumably different parent populations, e.g.
SNe~Ia in spiral galaxies and elliptical galaxies, which might be from
slightly different progenitor systems, but it is too early to draw
conclusions. The key to solving this question lies with observations
of local SNe~Ia. These objects can be observed with sufficient detail
that we can explore the different explosion models and possible
progenitor channels, which lead to the explosions.

\section{Conclusions}
\label{sec:conc}

Supernovae have been one of the main reasons, why we now consider a
dark energy component for the universe. These explosive events have
proved to be ideally suited for cosmological distance measurements.
Their variability, often regarded as detrimental by placing severe
observational constraints, has turned into an advantage. The
brightness evolution allows us to identify these cosmic light houses,
and, with sufficient knowledge of their intrinsic properties, we can
correct for various astrophysical effects, which could compromise the
cosmological deductions. 

Understanding the physics of the explosions remains a prime task.
Core-collapse supernovae have a relatively simple radiation transport
and can be used to derive fairly accurate distances in the local
universe. Core-collapse supernovae are fascinating events, which also
tell us about the stellar evolution of massive stars, how they shape
their environments through winds and how companions can change their
surface evolution, while the stellar core evolves towards the
collapse. Since at least some $\gamma-$ray bursts also show signatures
of supernovae, it is important to understand this supernova class
better. Through a modified expanding photosphere method they will
continue to provide further constraints on the Hubble constant. The
physical nature of this measurement is very attractive as it bypasses
the usual distance ladder. By expanding to higher redshifts an
independent confirmation of the accelerated expansion will become
possible. This method is observationally and theoretically expensive
requiring multi-band photometry and spectroscopy at several epochs and
tailored simulations of the spectra to match the observations.
Nevertheless, the effort should be continued as it appears at the
moment to be the only distance measurement to individual events to
complement the thermonuclear supernovae. 

Thermonuclear supernovae have spectacularly changed our view of the
universe. Empirically calibrated they have proved to be excellent
distance indicators. The fact that many questions regarding the exact
explosion mechanism and the as yet uncertain progenitor systems remain
has not hindered their use for cosmology. There is significant
progress in both areas. At the moment a consensus on these questions,
however, still remains to be found. The past decade has seen SNe~Ia take
centre stage for the derivation of cosmological parameters. While they
have been a favourite for the determination of the Hubble constant for
several decades, the difficult calibration of their absolute
luminosity at maximum has hampered their ability to determine an
accurate value free of systematics. With the exquisite capability of
SNe~Ia to deliver relative distances the problem of the Hubble
constant rests with an accurate calibration through other distance
indicators, e.g. Cepheid stars.

A beautiful confirmation of general relativity as the basis for the
cosmological model is the demonstration of time dilation in the light
curves and the spectral evolution of SNe~Ia. SNe~Ia discovered the
accelerated cosmic expansion and hence provided support for an
additional energy component of the universe. They now supply strong
evidence for a cosmological constant. The most recent supernova
surveys, based on over one hundred events, have not shown any
significant deviations from an integrated equation of state parameter
$\omega$=$-1$. One should, however, caution against any attempts to
over-interpret the current data. Exploring a time-variable $\omega$
should be done with the current limitations of the data in mind. The
accuracy required to significantly constrain $\omega(t)$ is probably
beyond what is currently available. 

Several systematic effects are still of concern for this
determination. The statistical uncertainties have reached the level of
these systematics and simply increasing the sample size beyond what
will be become available through SNLS and ESSENCE (several hundred
SNe~Ia beyond z$>$0.3) will not improve on the result any longer.
Detailed understanding of the various astrophysical effects, which
have to be treated to extract the cosmological signal, has now become
imperative. The physical nature of the light curve shape {\it vs.}
peak luminosity relation, the intrinsic colour variations among
SNe~Ia, the influence of dust absorption in the host galaxies,
evolutionary trends in SNe~Ia as a function of redshift and the
selection biases of the searches need to be examined carefully. The
limitations in accurately determining the local expansion rate are now
also becoming a significant weakness. The latter is an obvious
demonstration of the importance of the local SNe~Ia. They provide the
zero-point against which the distant supernovae are compared for the
cosmology. 

It is hence clear that an improved local sample of SNe~Ia will provide
several avenues for future improvements on the determination of the
dark energy parameters. In addition, supernovae projects extending to
higher redshifts and into the infrared hold great promise to overcome
the systematic problems encountered at the moment. 

Supernovae are one of the prime candidates to describe the
characteristics of dark energy. With the lack of a clear theoretical
contender for this unknown component, observations exploring the
effects of dark energy are decisive and hopefully will lead us
eventually to understand the properties of dark energy.

{\bf Acknowledgements} I am grateful for continuous discussions with
colleagues of the High-z Supernova Search Team, the Higher-z Team and
the ESSENCE Team as well as with team members of the Supernova
Cosmology Project and the Supernova Legacy Survey. This research was
supported by the DFG cluster of excellence 'Origin and Structure of the
Universe' (www.universe-cluster.de) and the DFG TransRegio TR33 'The Dark
Universe.'



\end{document}